# Hybrid-Supervised Deep Learning for Domain Transfer 3D Protoacoustic Image Reconstruction

## Yankun Lang[1], Zhuoran Jiang[2], Leshan Sun[3], Liangzhong Xiang [3] and Lei Ren[1]

[1] Department of Radiation Oncology Physics, University of Maryland, Baltimore, Baltimore, MD 21201, USA.

[2] Department of Radiation Oncology, Duke University, Durham, NC 27710, USA.

[3] Department of Biomedical Engineering and Radiology, University of California, Irvine, Irnive, California, 92617, USA.

E-mail: lren@som.umaryland.edu

**Abstract.** Protoacoustic imaging showed great promise in providing real-time 3D dose verification of proton therapy. However, the limited acquisition angle in protoacoustic imaging induces severe artifacts, which significantly impairs its accuracy for dose verification. In this study, we developed a deep learning method with a Recon-Enhance two-stage strategy for protoacoustic imaging to address the limited view issue. Specifically, in the Recon-stage, a transformer-based network was developed to reconstruct initial pressure maps from radiofrequency signals. The network is trained in a hybrid-supervised approach, where it is first trained using supervision by the iteratively reconstructed pressure map and then fine-tuned using transfer learning and self-supervision based on the data fidelity constraint. In the Enhance-stage, a 3D U-net is applied to further enhance the image quality with supervision from the ground truth pressure map. The final protoacoustic images are then converted to dose for proton verification. The results evaluated on a dataset of 126 prostate cancer patients achieved an average RMSE of 0.0292, and an average SSIM of 0.9618, significantly out-performing related start-of-the-art methods. Qualitative results also demonstrated that our approach addressed the limit-view issue with more details reconstructed. Dose verification achieved an average RMSE of 0.018, and an average SSIM of 0.9891. Gamma index evaluation demonstrated a high agreement (94.7% and 95.7% for 1%/3 mm and 1%/5 mm) between the predicted and the ground truth dose maps. Notably, the processing time was reduced to 6 seconds, demonstrating its feasibility for online 3D dose verification for prostate proton therapy.

## 1. Introduction

Proton therapy is a radiation treatment where proton beams are delivered to the target to disrupt and destroy tumor cells. After the protons enter the patient's body, the absorbed dose increases gradually at the beginning and then substantially at the end of the proton travel path, reaching a peak called Bragg Peak (BP), before dropping off sharply. This finite range and sharp dose falloff at the distal end of the BP increase our ability to conform radiation therapy treatment dose to the tumor and minimize collateral damage to neighboring critical organs. However, the precision of proton therapy is highly affected by the variations of patient positioning, anatomic structures, and dose calculation errors due to the sharp dose falloff of Bragg Peak. A small delivery error could cause a significant underdose to the target and an overdose to the healthy tissues. Therefore, online 3D dose verification during treatment is highly desirable in proton therapy to verify and minimize dose delivery errors to maximize its efficacy.

Over the years, many in-vivo dose verification methods have been developed to address this clinical need. Methods were developed to verify the proton dose range by measuring the dose or fluence with wireless implantable dosimeters [1, 3, 2], or by delivering and imaging separate proton beams [4, 5, 6]. However, these methods are not capable of fully verifying the tumor and organ at risk (OAR) dose since they don't provide the 3D volumetric information. Proton dose deposition can also be verified by measuring the surrogate data generated by proton irradiation. For example, Positron emission tomography (PET) [8, 7, 10, 9], prompt gamma (PG) imaging [11, 12, 13, 14] detects the gamma rays generated by irradiation along the proton beam path. Yuan et al in [15] used Magnetic resonance imaging to detect the radiobiological change of liver tissue after radiation. Specifically, MRI images were registered to the planning CT images. Then MR signal intensity (SI) was correlated to the radiation dose. Finally, dose–SI correlation was employed on registered MR images to estimate the proton end-of-range. These methods either don't provide online dose verification during treatment (PET, MRI) or have limited accuracy due to the low signal intensity and lack of 3D volumetric information (prompt gamma).

In recent years, protoacoustic (PA) imaging has been developed to detect proton-induced radiofrequency (RF) signals for dose verification[38, 39, 40, 41]. Specifically, the proton beam creates heat during the dose deposition, causing tissue expansion and contraction to generate acoustic waves, which can be detected by ultrasound transducers and used to reconstruct the pressure map and derive the corresponding dose deposition. Many researchers have conducted simulations on 2D Computed tomography (CT) images to verify dose range with protoacoustic signals [16, 17, 18]. More recently, matrix array transducers [19, 20] have been utilized for 3D ultrasound imaging, which showed a potential to provide real 3D online dose verification. The initial pressure map is reconstructed from the RF signals, and then related to the dose deposition. Traditional algorithms of reconstruction from the signal domain has been proposed. For example, Universal Back Projection (UBP) [23] projects the quantity calculated from the transducer measurements backward on a spherical surface within a solid angle, which is integrated to obtain the pressure with respect to position. This method suffers from distortion due to that tissue heterogeneity was not considered. Time Reveral (TR) [37, 22] is a method that iteratively updates the current pressure by adding the residual errors calculated with time reversed back-projection. Despite the progress, the reconstructed PA pressure map still suffers from severe distortion and artifacts due to the limited-angle view of the matrix array detector, limiting its accuracy for dose verification.

Deep learning-based methods have been developed in recent years to improve image reconstruction [24, 30, 31, 32]. Zhu et al. in [27] proposed a network that performs image reconstruction from RF signals directly by mapping the dual domain (signal-to-image) correlations with fully connected (FC) layers. Then they used a series of convolutional layers to denoise the output. However, this method is limited to memory capacity when dealing with high-resolution protoacoustic images. Häggström et al. proposed encoder-decoder architecture called DeepPET for direct PET image reconstruction [28]. To reduce memory consumption, DeepPET used convolutional layers rather than FC layers to learn a latent space representing the dual domain correlations. The latent space was then upsampled in the decoder to restore the image. However, this method ignored the consistency in the signal domain without accounting for the data fidelity constraint. Zhang el at. in [29] proposed a self-supervised learning method for ultrasound image reconstruction. The model is trained based on the data fidelity constraint, which minimizes the difference between the sinogram projected from the reconstructed image and the initially measured sinogram. Although this method has demonstrated improved reconstruction accuracy in ultrasound images, it still suffers from severe distortion artifacts when applied for protoacoustic images due to the limited angle view issue. To address limitations in image reconstruction, deep learning has also been developed for post-reconstruction image enhancement [25]. To address the issue in limited view PA reconstruction, Jiang et al. [26] utilized a 3D U-net that enhances an initial pressure map reconstructed by TR to reduce the distortion artifact, then derived a 3D dose map for dose verification. Despite the improvements, the efficacy of deep learning enhancement is limited by the quality of the initial reconstruction. The initial pressure map reconstructed by the TR method suffers from severe distortion with many detailed anatomical structures lost, which consequently impairs the accuracy of the image enhancement afterward. Meanwhile, this method suffers from time consuming in testing stage since TR method needs numerous time (120 s) for reconstruction. The low efficiency makes this method impractical for online dose verification. Recently, transformer network has been applied in various medical imaging research [33, 34] due to its long-range dependency and adaptive self-attention characteristics. Swin Transform [35] was proposed with moving receptive field windows of reduced size to greatly reduce the computational complexity. Huang el at. [36] utilized a Swin transformer-based generator to enhance the quality of k-space downsampled MRI images. A discriminator was used to distinguish the enhanced result from ground truth to improve the accuracy further. This work demonstrated that the transformer-based models showed great performance in enhancing MRI image quality after reconstruction.

To address the limited angle view problem in PA imaging and further improving the reconstruction quality, in this study, we proposed a deep learning-based protoacoustic image reconstruction method, where a Recon-Enhance two-stage strategy is applied as shown in Fig.1 to harness the power of deep learning for both image reconstruction and post-reconstruction enhancement. Specifically, in the Recon-stage, the proposed network directly reconstructs the image from RF signals with hybrid supervision and transfer learning. In the Enhance-stage, a 3D U-net is applied to further improve the image quality. Compared with the method in [26], where the reconstruction was implemented by Time Reversal, our approach directly reconstructs the initial pressure map from raw RF signals, which can significantly reduce the processing time and improve the accuracy since more essential structural information can be preserved. The main contributions of our article are multi-fold: 1) An end-to-end image reconstruction and enhancement strategy using deep learning is developed for PA imaging to significantly improve its quality; 2) We apply convolutional layers rather than fully connected layers to construct a domain-transfer module to address the memory consumption problem, while maintaining a higher inference speed; 3) We replace the general convolutional layers with transformers to build our network for its long-range dependency, and proposed a novel hybrid supervison method to keep the data fidelity consistency; 4) The proposed method is evaluated on protoacoustic data generated from the CT images and clinical treatment plans of prostate cancer patients, demonstrating the feasibility of high precision 3D dose verification in proton therapy.

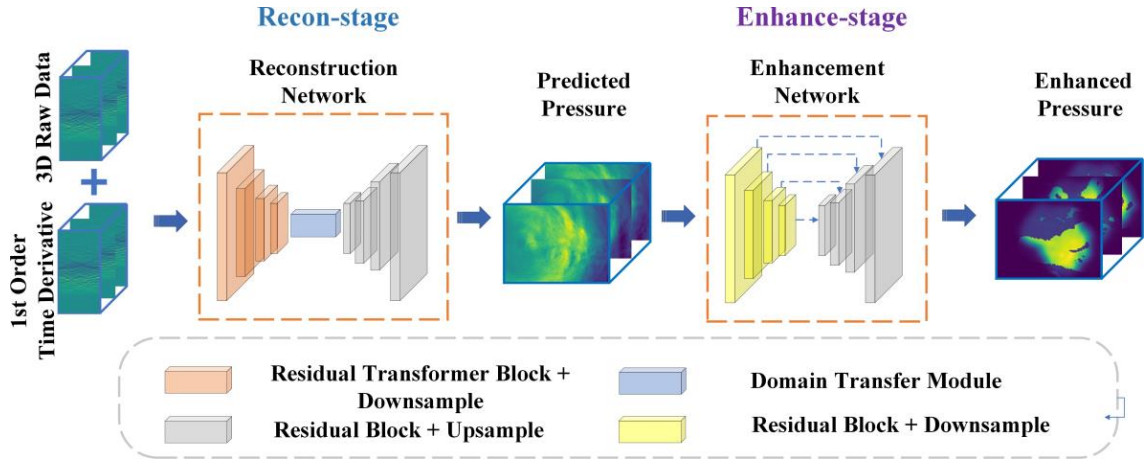

**Figure 1.** The workflow of our approach for protoacoustic image reconstruction with a Recon-Enhance strategy.

## 2. Methods

### 2.1. Problem Formulation

During protoacoustic process, proton deposits energy when traveling through the patient's body, causing tissue temperature to rise and generating acoustic signals, which can be formulated as:

$$(\nabla^2 - \frac{1}{c^2}\frac{\partial^2}{\partial t^2})p(\boldsymbol{r}, t) = -\frac{\Gamma}{c^2}H(\boldsymbol{r})\frac{\partial \delta(t)}{\partial} \qquad (1)$$

where $p(\boldsymbol{r}, t)$ denotes the measured pressure at location $\boldsymbol{r}$ at time $t$. $H(\boldsymbol{r})$ denotes the initial pressure. $c$ is speed of sound in the medium. $\Gamma$ is the dimension less Gruneisen parameter, and $\delta(t)$ denotes the delta function. The objective of our study is to reconstruct the initial map $H(\boldsymbol{r})$ from the measurements $p(\boldsymbol{r}, t)$.

Universal backprojection (UBP) [23] is a linear reconstruction method derived from equation (1), which can be formulated as:

$$H(\text{r}) = \sum_{i=1}^{N} b(\boldsymbol{d}_i, t) = \frac{|r - \boldsymbol{d}_i|}{c} \frac{\Delta\Omega_i}{\sum_{i=1}^{N} \Delta\Omega_i} \qquad (2)$$

where $\boldsymbol{d}_i$ and $\Delta\Omega_i$ denote the position and solid angle, respectively. $b(\boldsymbol{d}_i, t)$ is the backprojection term of the $i$-th transducer, which can be formulated as:

$$b(\boldsymbol{d}_i, t) = 2p(\boldsymbol{d}_i, t) - 2t \frac{\partial p(\boldsymbol{d}_i, t)}{\partial t} \qquad (3)$$

Apparently, the reconstruction of the initial pressure map $H(\boldsymbol{r})$ from the measurements $p(\boldsymbol{r}, t)$ critically depends on the first-order partial derivative $\partial p(\boldsymbol{d}_i, t)/\partial t$, which can be used as prior knowledge for our model design.

Direct reconstruction of high-quality initial pressure map from RF signals is challenging since the network needs to balance the domain transfer for image reconstruction and the enhancement to correct the distortions caused by limited view in PA images. To address this problem, a recon-enhance strategy is proposed, as shown in Fig.1, to first use a network for image reconstruction to generate an initial pressure map with reasonable quality. Then another network is applied afterward to further enhance the reconstructed images.

## 2.2. Domain Transfer Reconstruction Network (DTR-Net)

The overview of the proposed Domain Transfer Reconstruction Network (DTR-Net) is shown in Fig.2(a). DTR-Net utilizes a contracting-expanding architecture, taking both the 3D RF image $\boldsymbol{S} \in R^{H^s \times W^s \times D^s}$ and the corresponding first order derivative image $\partial \boldsymbol{S}/\partial \boldsymbol{t} \in R^{H^s \times W^s \times D^s}$ as input, where $H^s$, $W^s$ and $D^s$ represent the height, width and depth of the RF image, respectively. The contracting path consists of four residual transformer blocks (RTBs) followed by down-sampling layer to extract high level features as shown in Fig.2(a). Each RTB shown in Fig.2(b) is built by several 3D Swin transformer (ST) layers shown in Fig.2(c) due to its characteristic of long-range dependency. Swin transformer [35] was developed from the original Transformer layer where window based multi-head self-attention (W-MSA) is implemented. Specifically, given a feature map, the ST layer first partitions the input into several non-overlapping windows. For each local window feature $F$, the query ($Q$), key ($K$) and value ($V$) matrices are calculated by:

$$Q = FP_Q, K = FP_K, V = FP_V, \qquad (4)$$

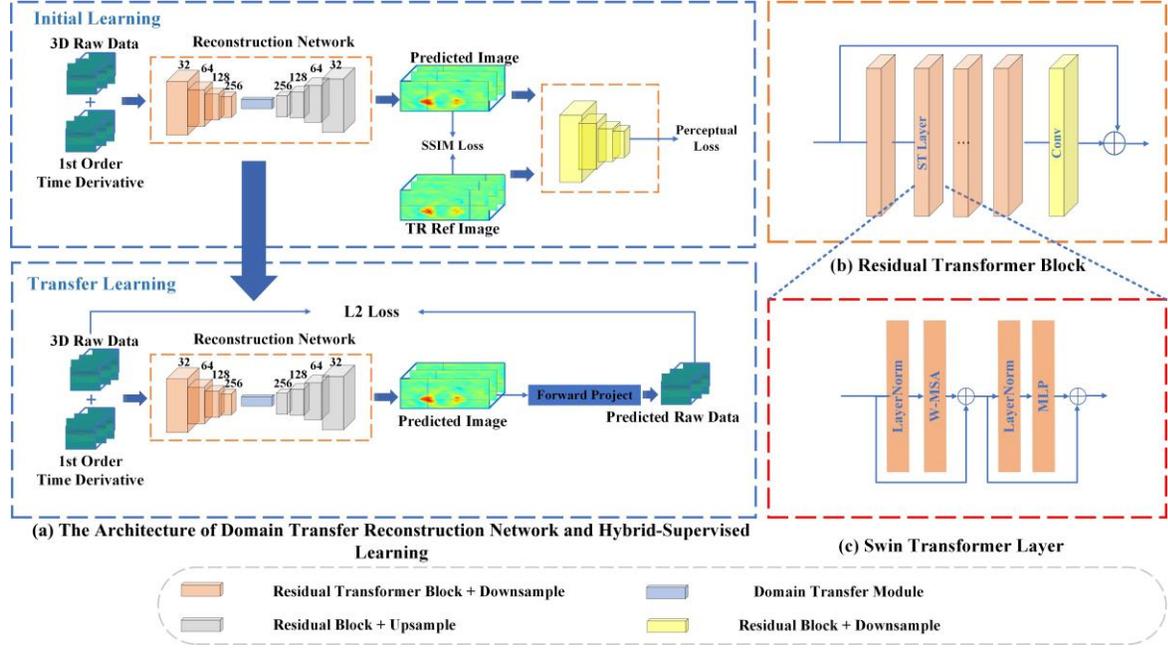

**Figure 2.** The architecture of the proposed DTR network. (a) Two-step hybrid-supervised training where the network learning weights are shared across the initial learning and transfer learning steps. (b) Details of the residual transformer block. (c) Details of the swin transformer layer.

where $P_Q$, $P_K$ and $P_V$ are the projection matrices. Then, a self-attention mechanism is applied to calculate the attention matrix by:

$$Att(Q, K, V) = \sigma(QK^T / + B)V, \tag{5}$$

where $B$ is the relative positional encoding. $\sigma$ denotes the softmax activation fuction. The final output of the ST layer is computed as:

$$F_{att} = WMSA(Norm(F)) + F, \tag{6}$$

$$F_{ST} = MLP(Norm(F_{att})) + F_{att} \tag{7}$$

where $Norm$ denotes layer normalization. $MLP$ denotes multi-layer perceptron with two fully connected layers for further feature transformations. Residual connection is applied here for feature consistency as shown in Fig.2(c). We applied 2, 4, 8 and 16 STs in each TB respectively to extract hierarchy features. The final extracted feature map $F^S \in R^{\frac{H^s}{16} \times \frac{W^s}{16} \times \frac{D^s}{16}}$ is fed into a domain transfer module, which is simply built by a learnable convolution layer, resizing the feature map from $\frac{H^s}{16} \times \frac{W^s}{16} \times \frac{D^s}{16}$ to $\frac{H^i}{16} \times \frac{W^i}{16} \times \frac{D^i}{16}$, where $H^i$, $W^i$ and $D^i$ represent the height, width and depth of the initial pressure map, respectively. The expanding path consists of 4 residual blocks, each of them is built by a up-sampling layer, and two consistent 3D convolution layers with $3 \times 3 \times 3$ kernel, followed by ReLU activation and Group normalization layers. Finally, a convolution layer with a kernal size $1 \times 1 \times 1$ is applied to output the reconstructed initial pressure map $P \in R^{H^i \times W^i \times D^i}$. Notably, skip connection is not applied due to the domain inconsistency of the feature maps between the contracting and expanding paths.

The reconstruction network is trained using hybrid supervision with transfer learning, as explained below:

### 2.2.1. Initial Training

As shown in Fig.2(a), in the initial training, the model is trained to reconstruct PA images by minimizing the difference between the reconstructed pressure map $P$ by the model and the reference pressure map $P^*$ reconstructed by the TR method. Since iterative TR can recover most of the reconstruction details, we utilize the TR results as the reference for initial training. Both SSIM loss $L_{ssim}(P, P^*)$ and perceptual loss $L_{perc}$ are used to enhance the stability of the reconstruction. A perceptual loss calculates the difference between features yielded by a designed VGG network. The training loss is:

$$L_I = \alpha_1 L_{ssim}(P, P^*) + \beta_1 L_{perc}(P, P^*), \tag{8}$$

where $\alpha_1 = 1.0$ and $\beta_1 = 0.025$ are the training weights that have been set empirically. This step enables DTR-Net focus on discovering the most representative features for fast reconstruction.

### 2.2.2. Transfer Learning

Considering that the inverse problem is ill-posed and the TR reconstruction is prone to artifacts itself, we applied self-supervised transfer learning to further improve the reconstruction network based on data fidelity constraint. Specifically, as shown in Fig.2(a), the network is fine-tuned using transfer learning and self-supervision based on data fidelity constraint, which forces the projected RF data from the reconstructed images to match the measured raw data. The forward projection of RF data from the reconstructed images is carried out using Matlab k-wave toolbox [37]. The loss function $L_{TL}$ used for transfer learning (TL) is defined as a $l2$ loss to evaluate the difference between the the input RF signal $S$ and the predicted RF data $S^*$ in the data fidelity constraint, as shown below:

$$L_{TL} = \sum_{i=1}^{N}(S_i - S_i^*)^2 \tag{9}$$

where $N$ denotes the entire number of image voxels. This step enables the network to further fine-tune the reconstruction solely based on the raw data, thus removing the impact of imperfect supervision by the TR reconstructed images in the initial training to improve the reconstruction quality.

## 2.3. Enhancement and Dose Conversion Network

Due to the limited angle scan of PA imaging, the image generated by the reconstruction network can still have residual artifacts, such as image distortion. A 3D U-Net will be applied to further enhance the reconstructed images to address the residual artifacts. The network has the same architecture and parameter setting as proposed in [26]. Specifically, the network takes the reconstructed results as input, and output the residual difference between the input and ground truth. Model weights are optimized by minimizing the mean squared error (MSE) loss between enhanced and ground truth images during the training. The final result is obtained by adding the output of the enhancement network to the input.

Finally, the enhanced pressure map is converted to dose map for proton dose verification. Specifically, an initial dose map was calculated by dividing the reconstructed pressure map by the dose conversion coefficient map derived from patient CT images. A 3D U-net was developed with the same architecture and training settings as in [26] to predict the residual errors compared with the ground truth to generate the final dose map.

## 2.4. Training Implementation and Inference

The models in both Recon-Enhance stages were trained by an ADAM optimizer with an initial learning rate of 0.001, reduced by a factor of 5 after every $500,000$ epochs. In the Recon-stage, we set $\alpha_1 = 1.0$ and $\beta_1 = 0.025$ for initial training with the loss defined by equation (10). After $3,000,000$ epochs, we started the transfer learning with the loss defined in equation (9) for another $1,000,000$ epochs. Finally, in the Enhance-stage, we train the enhancement network for another $1,000,000$ epochs. The entire training process takes about 3 days to be finished.

During the inference, the trained DTR-Net uses RF data measured by limited angle PA imaging to reconstruct the pressure map, which is then enhanced by the enhancement network to generate the final PA images. This recon-enhance approach takes less than 6 second to process a 3D RF signal image with the size of $32 \times 32 \times 112$ to reconstruct a PA pressure map with the size of $48 \times 48 \times 112$. The network was implemented based on Pytorch with a 40 GB Nvidia server GPU and a 64 GB RAM.

## 2.5. Data Collection

In this study, a dataset consisting of 126 anonymized patients with prostate cancers was collected under an IRB approved protocol. Data of each patient contains the planning CT scan and the corresponding clinical treatment plan. Dose map of the plan was provided by a commercial software named RayStation (RaySearch Laboratories, Stockholm/Sweden), and then normalized to the maximum dose. Each CT scan was firstly segmented into four categories: air, fat, soft tissue and bone according to the predefined HU value thresholding. All the tissue-specific parameters including the density, sound of speed, and the Gruneisen parameter are predefined in Table 1.

The acoustic simulation for generating the RF signal started with the calculation of the initial pressure (P0) by multiplying the dose map with the tissue density and the Grüneisen parameter:

$$P0 = dose\_map \times \rho \times \Gamma, \tag{10}$$

Then, the simulation was performed using the open-source k-wave toolbox on Matlab. Specifically, a planar detector of 8 cm×8 cm with a 64 × 64 ultrasound transducer array was simulated below the prostate and near the perineum area with a $\frac{\pi}{6}$ tilt angle to cover the prostate area and avoid the pelvic bones. The central frequency of each transducer element was set to 500kHz with 100 bandwidth and a sampling rate of 5MHz. Tissue-specific heterogeneity and attenuation were considered during the acoustic signal propagation. Finally, a Gaussian white noise with 10dB signal-to-noise ratio (SNR) was added to the acquired RF signals, which is used as input of our network. TR method was applied for 10 iterations to reconstruct the initial pressure maps from the simulated RF signals, which are used as ground truth for the initial training of DTR-Net in the Recon-stage in Fig.2(a). The initial pressure map P0 and the dose map were used as the ground truth for training the pressure map enhancement network and the dose conversion network. Both the pressure map and dose map were resampled to the resolution of $2.50 \times 2.50 \times 1.25 mm^3$ with the size of 48 × 48 × 112, and the simulated RF signal was resampled to the size of 32 × 32 × 112 to reduce the memory consumption.

**Table 1.** Tissue-specific parameter setting for RF signal simulation. $v$, $\rho$ and $\Gamma$ refer to the speed of sound, tissue density and the Grüneisen parameter, respectively. $\alpha$ denotes the attenuation coefficient.

| Tissue | HU value | $v$ (m/s) | $\rho$ (kg/$m^3$) | $\Gamma$ | $\rho \times \Gamma$ (kg/$m^3$) | $\alpha$ (dB/cm/MHz) |
|---|---|---|---|---|---|---|
| Air | [-1000,-200) | - | - | - | - | - |
| Water | air overwritten | 1500 | 1000 | 0.11 | 110 | 0.0022 |
| Fat | [-200,-50) | 1480 | 920 | 0.80 | 736 | 0.5 |
| Soft tissue | [-50,100) | 1540 | 1040 | 0.30 | 312 | 1 |
| Bone | [100, max) | 2000 | 1900 | 0.80 | 1520 | 10 |

## 3. Experiments and Results

### 3.1. Data Augmentation

We perform data augmentation to improve the model generalization while avoiding over-fitting. In this study, the PA detector was simulated at different positions in the perineum area to generate more RF-P0 pairs to enlarge the training set. Specifically, the detector was located below the prostate and near the perineum area with an initial $\frac{\pi}{6}$ degree tilt angle. Then the detector was rotated along the lateral axis by different angles that are equally sampled within a range of $[-\frac{\pi}{6}, \frac{\pi}{6}]$ that covers the whole prostate area, as shown in Fig.3. For each sampled angle, protoacoustic simulation procedures was performed to generate the corresponding RF signals from the initial pressure map P0. The augmentation was repeated for 20 times with equally spaced angles for each patient. Using 5-fold cross-validation, we randomly selected 66 patients for training, 20 patients for validation, and the rest 40 patients for testing.

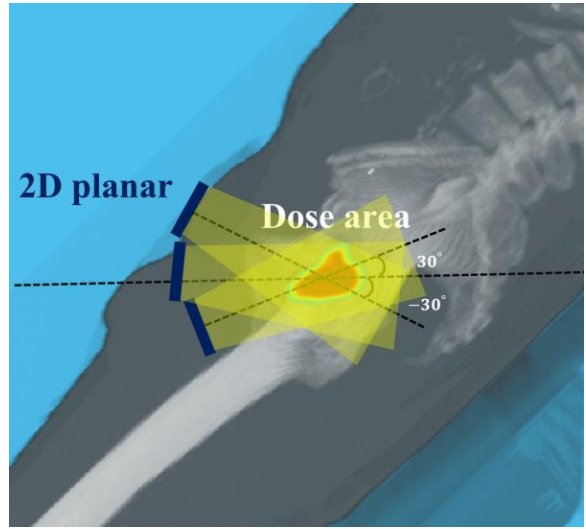

**Figure 3.** Augmentation setup. To simplify the visualization, we fix the patient body and rotate the 2D plannar matrix along the lateral direction to cover the entire prostate that is illustrated by the dose area. Three examples of acquisition are given in this figure.

### 3.2. Competing Methods

We quantitatively and qualitatively compared our method with three baseline methods:

- **Time Reversal**: An iterative method for image reconstruction. In each iteration, a pressure map is reconstructed based on forward projection, then the time parameter is reversed and a RF signal is calculated from the reconstructed pressure map and compared with the acquired RF signals. The current pressure is updated by adding a residual pressure obtained by back-projecting the RF signal differences. In this experiment, TR method was repeated for 10 iterations empirically considering the balance between reconstruction quality and time consumption to reconstruct the initial map.

- **Method in [26]**: A state-of-the-art deep-learning method that jointly performs initial pressure reconstruction and dose verification. The first network takes the pressure reconstructed by TR method as input, and outputs a result with enhanced quality. An initial dose map is generated by multiplying the reconstructed results with the dose coefficients derived from the CT scans, and then further refined by the second network. We trained the network for pressure reconstruction using the same architecture and paremeter setting as described in [26] with the input size of 48×48×112. Same training/validation set and augmentation method were applied for the training.

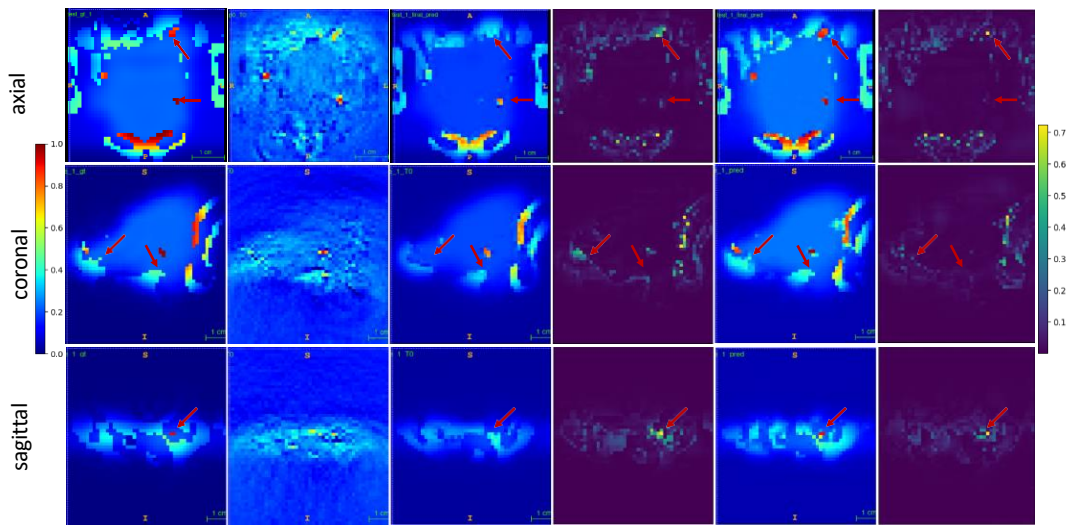

**Figure 4.** Example initial pressure reconstruction results (Normalized). From left to right: Colorbar of the initial maps, ground truth, results by using TR, results by using the method in [26], difference maps between ground truth and results from [26], results by our approach, and difference maps between ground truty and our results, colorbar of the difference maps.

**Table 2.** Quantitative analysis of the reconstruction results of initial pressure maps and dose verification

| Modality | Method | MSE | PSNR (dB) | SSIM | Speed (s) |
|---|---|---|---|---|---|
| PA Image | Time Reversal | 0.145 ± 0.059 | 24.02 ± 0.58 | 0.854 ± 0.045 | 120 |
| | Method in [26] | 0.033 ± 0.021 | 29.6 ± 0.34 | 0.939 ± 0.013 | 120 |
| | DTR-A | 0.042 ± 0.029 | 26.52 ± 0.41 | 0.892 ± 0.015 | 6 |
| | DTR-B | 0.030 ± 0.014 | 30.21 ± 0.37 | 0.959 ± 0.015 | 6 |
| | Our approach | **0.029 ± 0.011** | **30.37 ± 0.26** | **0.962 ± 0.013** | 6 |
| Dose Verification | Method in [26] | 0.026 ± 0.013 | 31.79 ± 0.34 | 0.973 ± 0.016 | 120 |
| | Our approach | **0.018 ± 0.009** | **34.86 ± 0.27** | **0.989 ± 0.007** | **6** |

### 3.3. Pressure Map Reconstruction Results

The reconstruction quality was evaluated by comparing the predicted pressure with the ground truth using root mean squared errors (RMSE). Additionally, we also compared Peak signal-to-noise ratio (PSNR) and structural similarity index measure (SSIM) to further investigate the performance on details and basic structure reconstruction. The overall quantitative results of pressure map reconstruction are summarized in Table 2. The qualitative results are also shown in Fig.4. Among the three compared methods, TR method results in the largest RMSE (0.145) and the lowest SSIM (0.854). The method in [26] significantly improved the reconstruction quality by reducing the RMSE to 0.033. Meanwhile, the SSIM was improved to 0.939, demonstrating the effectiveness of using 3D U-net for quality enhancement. However, details were still not reconstructed

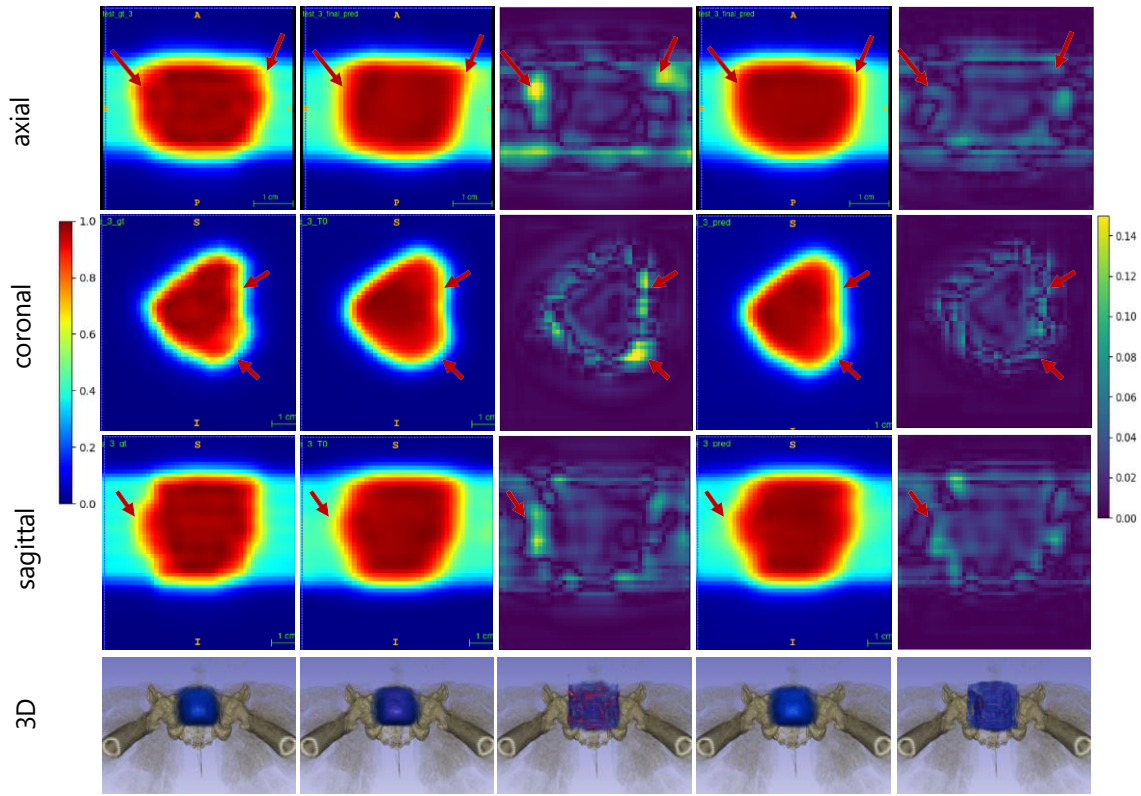

**Figure 5.** Example dose verification results (Normalized). From left to right: Color bar of the dose maps, ground truth, results by using the method in [26], difference maps between ground truth and results from [26], results by our approach, difference maps between ground truth an our results, and colorbar of the difference maps.

in some challenging locations, while the whole structure was blurred.

Our method is more accurate than all compared methods, with a RMSE error as low as 0.029. As shown in Fig.4, most of the details were successfully reconstructed in the challenging areas while the blur effect was eliminated, suggesting the effectiveness of the explicit learning of correlation between the image and signal domains. Specifically, the SSIM was improved to 0.962, showing a high similarity of anatomic structure compared with the ground truth, confirming the effectiveness of using SSIM and perceptual losses for training. RMSE and SSIM results are boxplotted in Fig.6(a) and (b) (see supplementary material). Notable, we also compared the runtime for testing using different methods. TR and method in [26] both took about 2 minutes to process a single case due to iterations. Our approach achieved the fastest speed taking as low as 6 seconds, making the method much more applicable for online dose verification in proton therapy.

### 3.4. Dose Verification Results

We compared the dose maps that were predicted from the pressure maps recontructed by our approach and the method in [26], in terms of RMSE, PSNR and SSIM. Table 2 also gives the quantitative results, where our approach gains significant improvements. Particularly, our approach reduces the RMSE from 0.026 to 0.018, and increases the SSIM from 0.973 to 0.989, showing a high similarity between the predicted and the ground truth 3D dose maps. Fig.5 shows qualitative results of several challenging cases, where the dose maps restored by using our method show more accuracy, due to the high quality of the input pressure maps. Finally, the pressure reconstruction and dose prediction with the proposed method only take about 6 seconds in total.

**Table 3.** Quantitative analysis of the reconstruction results of dose maps.

| Modality | Metric | Method in [26] | Our approach |
|---|---|---|---|
| Dose | Gamma Index (3%/3mm) | 97.9% ± 1.1% | **99.3% ± 0.4%** |
| | Gamma Index (3%/5mm) | 98.3% ± 0.8% | **99.6% ± 0.3%** |
| | Gamma Index (2%/3mm) | 95.7% ± 2.2% | **97.1% ± 1.9%** |
| | Gamma Index (2%/5mm) | 96.5% ± 2.0% | **97.8% ± 1.8%** |
| | Gamma Index (1%/3mm) | 92.7% ± 2.5% | **94.7% ± 2.5%** |
| | Gamma Index (1%/5mm) | 93.7% ± 2.4% | **95.7% ± 2.5%** |

Additionally, we compared the predicted dose maps with the ground truth in terms of gamma index as shown in Table 3 and Fig.6(c) (see supplementary material). Our approach increased the gamma index from 97.9% to 99.3%, from 98.3% to 99.6%, from 95.7% to 97.1%, and from 96.5% to 97.8% for 3%/3 mm, 3%/5 mm, 2%/3 mm, and 2%/5 mm, respectively. Notably, our approach achieved high gamma index rates as 94.7% and 95.7% for 1%/3 mm and 1%/5 mm, showing a high agreement between the predicted and the ground truth dose maps, which further demonstrates the effectiveness of our approach.

### 3.5. Ablation Study

We performed an ablation study by comparing our approach with two variants: 1) DTR-A, where transfer learning was not applied. We used the loss function defined in equation (10) to train the network in the Recon-stage, then performed enhancement in the Enhance-stage; 2) DTR-B, where we kept the same network architecture and training losses that were used in our proposed method, except that we used the initial pressure rather than TR results as the ground truth in the Recon-stage. The reconstruction results were quantitatively evaluated with RMSE, PSNR and SSIM. All compared methods were trained using the same augmented dataset.

The results of the ablation study, denoted as DTR-A and DTR-B, are also summarized in Tables 2 and Fig.6(a) and (b) (see supplementary material). Specfically, DTR-A had the highest RMSE (0.042) and the lowest SSIM (0.892). Compared with DTR-A, DTR-B significantly improves the reconstruction quality with a RMSE of 0.030 and SSIM of 0.959, confirming the effectiveness of transfer learning. Our approach achieved the lowest RMSE and the highest SSIM.

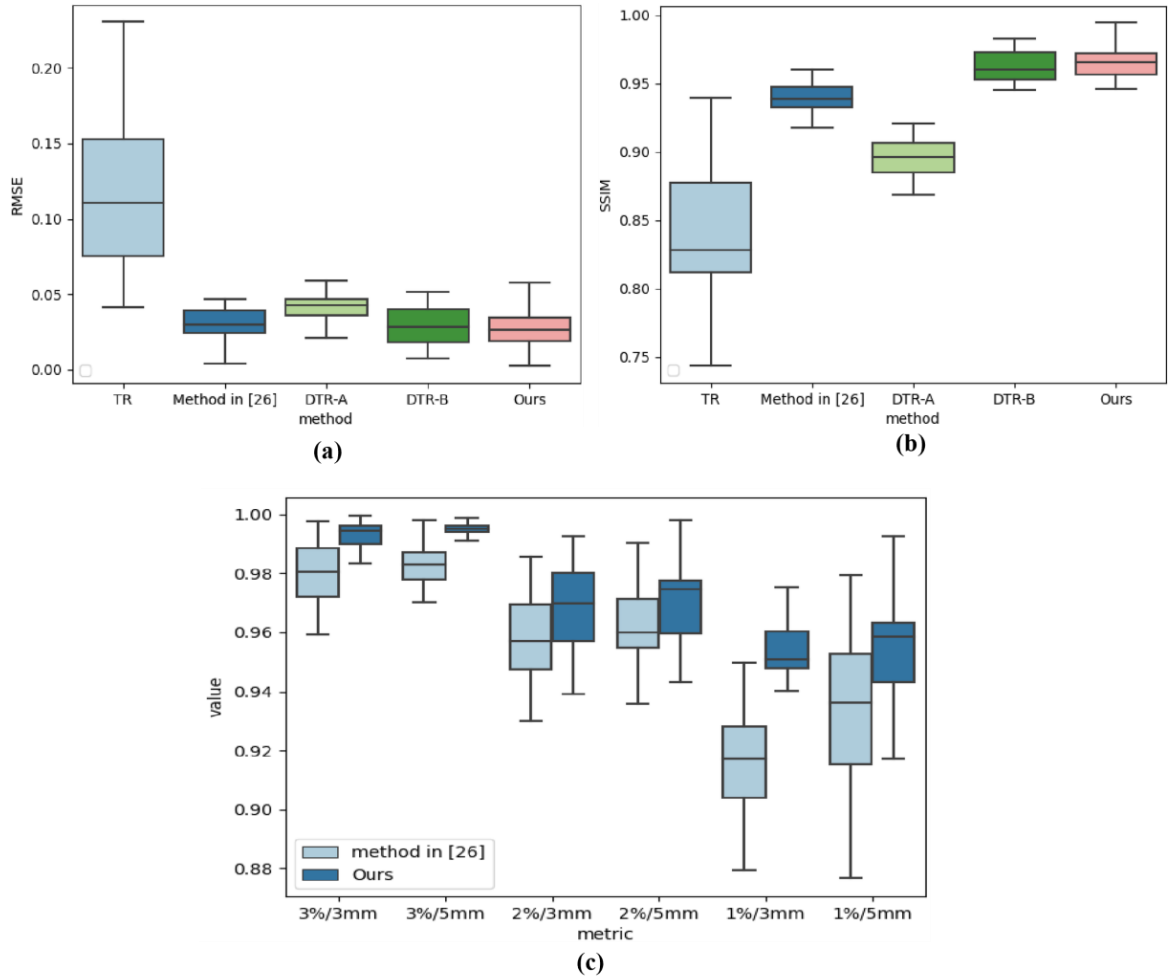

**Figure 6.** Boxplot of the (a) RMSE of the predicted pressure maps using different methods, (b) SSIM of the predicted presure maps using different methods, and (c) Gamma index of the predicted dose maps using different metrics.

## 4. Discussion

### 4.1. Pressure and Dose Reconstruction for Protoacoustic Imaging

Our approach used transformer-based blocks to build the network, which is trained by hybrid-supervision for reconstructing the initial pressure map directly from the raw RF signals. Results showed that our approach has gained a significantly improved accuracy and speed. For the compared TR method, due to the limited angle view of the 2D matrix array, the reconstructed pressure map suffers from severe distortions, where most of the structure details cannot be distinguished. The method in [26] applied a 3D U-net to enhance the quality of the initial map reconstructed by TR method. However, the efficacy of the network enhancement is limited by the quality of the TR reconstruction. Specifically, in areas where the TR image is severely distorted with missing details, image enhancement will not be able to recover anatomical details that are completely lost in the input image as shown in Fig.5 highlighted by red arrows. For DTR-A, without the transfer learning to tune the model based on data fidelity constraint, the reconstruction is highly affected by the limited quality of TR reconstruction used as the reference in the initial training, leading to suboptimal results. DTR-B significantly improved the accuracy. However, using initial pressure as the ground truth requires the network to perform both domain correlation learning for image reconstruction and correction of image distortion caused by the limited-angle acquisition, which is hard to balance during the training and leads to slightly lower quality compared with the proposed method. Increasing learning parameters could potentially solve this problem but will cause more memory consumption. Our approach directly reconstructed the pressure from RF signals to preserve the essential structural information. Using TR results for initial training in the Recon-stage made the network focus on domain transfer mapping, thus improved the training efficiency and efficacy. Meanwhile, the transfer learning with self-supervision based on data fidelity constraint ensured consistency in both domains. Fig.4 and Fig.5 showed that our approach can successfully reconstruct most structural details, leading to high quality 3d dose verification result. Quantitative results also demonstrated the superiority of our method compared to other methods. Another major advantage of the deep learning reconstruction network is its high efficiency. The proposed network achieved an end-to-end processing time of 6s, which is substantially shorter than the 2 min required by the TR method. This high efficiency is critical for the clinical adoption of the technique since time is of the essence when performing online dose verification during proton therapy.

## 4.2. Limitations and Future Work

There are some limitations of this study. First, since we applied k-wave toolBox to perform projection from the image domain back to the signal domain during the transfer learning, the training time has increased numerously. Second, the reconstruction quality is still expected to be improved, although our method has eliminated the distortion and artifacts caused by limited angle view.

In the future work, we will focus on developing a deep learning method to automatically learn the back-projection mapping to accelerate the training process. Besides, we will investigate RF signal pre-processing to improve the RF signal quality, which can further improve the performance of our proposed method. We will also apply our approach to other image modalities to verify the generalization of the proposed network.

Our approach was evaluated on simulated data due to the lack of patient experiments. Simulation data have the advantage of providing the ground truth of initial pressure and dose map for evaluation compared to real patient data where ground truth is often unavailable. The simulation parameters were set empirically to make the simulation results close to real data. Experimental and real patient studies are warranted in the future to further evaluate the clinical efficacy of the technique.

## 5. Conclusion

In this work, we have proposed a hybrid-supervised deep learning method to reconstruct PA images for proton therapy dose verification. DTR-Net using transformer blocks, transfer learning and hybrid supervision has been developed for direct PA image reconstruction from the raw RF signals, and image enhancement has been applied to solve the limited angle view problem. The results show that our method significantly outperforms competing state-of-the-art methods. Most importantly, our approach achieved superior performance on reconstructing 3D dose with a fast processing speed, making it very practical for online 3D dose verification in proton therapy.

### Acknowledgment

This work was sponsored in part by National Institutes of Health grants R01 EB032680 and R01 EB028324.

### References

[1] H. Lu, G. Mann, and E. Cascio, "Investigation of an implantable dosimeter for single-point water equivalent path length verification in proton therapy," *Medical physics*, vol.37, no.11, pp.5858-5866, 2010.

[2] E. H. Bentefour, T. Shikui, D. Prieels, and H. M. Lu, "Effect of tissue heterogeneity on an in vivo range verification technique for proton therapy," *Physics in Medicine & Biology*, vol.57, no.17, pp.5473, 2012.

[3] J. Telsemeyer, O. Jäkel, and M. Martišíková, "Quantitative carbon ion beam radiography and tomography with a flat-panel detector," *Physics in Medicine & Biology*, vol.57, no.23, pp.7957, 2012.

[4] U. Schneider, et al., "First proton radiography of an animal patient," *Medical physics*, vol.31, no.5, pp.1046-1051, 2004.

[5] U. Schneider, E. Pedroni, M. Hartmann, J. Besserer, and T. Lomax, "Spatial resolution of proton tomography: methods, initial phase space and object thickness," *Zeitschrift für Medizinische Physik*, vol.22, no.2 pp.100-108, 2012.

[6] S. N. Penfold, A. B. Rosenfeld, R. W. Schulte, and K. E. Schubert, "A more accurate reconstruction system matrix for quantitative proton computed tomography," *Medical physics*, vol.36, no.10, pp:4511-4518, 2009.

[7] F. Fiedler, et al., "In-beam PET measurements of biological half-lives of 12C irradiation induced β±activity," *Acta Oncologica*, vol.47, no.6, pp.1077-1086, 2008.

[8] F. Fiedler, et al., "On the effectiveness of ion range determination from in-beam PET data," *Physics in Medicine & Biology*, vol.55, no.7 pp:1989, 2010.

[9] T. Nishio, et al., "The development and clinical use of a beam ON-LINE PET system mounted on a rotating gantry port in proton therapy." *International Journal of Radiation Oncology\* Biology\* Physics*, vol.76, no.1, pp.277-286, 2010.

[10] A. Miyatake, et al., "Measurement and verification of positron emitter nuclei generated at each treatment site by target nuclear fragment reactions in proton therapy," *Medical physics*, vol.37, no.8, pp.4445-4455, 2010.

[11] C. H. Min, H. R. Lee, C. H. Kim, and S. B. Lee, "Development of array-type prompt gamma measurement system for in vivo range verification in proton therapy," *Medical physics*, vol.39, no.4, pp.2100-2107, 2012.

[12] J. C. Polf, S. Peterson, G. Ciangaru, M. Gillin, and S. Beddar, "Prompt gamma-ray emission from biological tissues during proton irradiation: a preliminary study," *Physics in Medicine & Biology,* vol.54, no.3, pp.731, 2009.

[13] E. Draeger, et al., "3D prompt gamma imaging for proton beam range verification," *Physics in Medicine & Biology,* vol.63, no.3, 035019, 2018.

[14] T. Kormoll, et al., "A Compton imager for in-vivo dosimetry of proton beams—A design study", *Nuclear Instruments and Methods in Physics Research Section A: Accelerators, Spectrometers, Detectors and Associated Equipment,* vol.626, pp.114-119, 2011.

[15] Y. Yuan, et al., "Feasibility study of in vivo MRI based dosimetric verification of proton end-of-range for liver cancer patients", *Radiotherapy and Oncology,* vol.106, no.3, pp.378-382, 2013.

[16] C. Freijo, J. L. Herraiz, D. Sanchez-Parcerisa, and J. M. Udias, "Dictionary-based protoacoustic dose map imaging for proton range verification", *Photoacoustics,* vol.21, 100240, 2021.

[17] S. Yao, Z. Hu, Q. Xie, Y. Yang, and H. Peng, "Further investigation of 3D dose verification in proton therapy utilizing acoustic signal, wavelet decomposition and machine learning", *Biomedical Physics & Engineering Express* vol.8, no.1, 015008, 2021.

[18] Y. Yu, Z. Li, D. Zhang, L. Xing and H. Peng, "Simulation studies of time reversal-based protoacoustic reconstruction for range and dose verification in proton therapy", *Medical physics,* vol.46, no.8, pp.3649-3662, 2019.

[19] J. Yu, H. Yoon, Y. M. Khalifa, and S. Y. Emelianov, "Design of a volumetric imaging sequence using a vantage-256 ultrasound research platform multiplexed with a 1024-element fully sampled matrix array", *IEEE transactions on ultrasonics, ferroelectrics, and frequency control,* vol.67, no.2, pp.248-257, 2019.

[20] M. Wang, et al., "Toward in vivo dosimetry for prostate radiotherapy with a transperineal ultrasound array: A simulation study", *IEEE transactions on radiation and plasma medical sciences*, vol.5, no.3, pp.373-382, 2020.

[21] B. E.Treeby, E. Z. Zhang, and B. T. Cox, "Photoacoustic tomography in absorbing acoustic media using time reversal", *Inverse Problems,* vol.26, no.11, 115003, 2010.

[22] Y. Hristova, P. Kuchment, and L. Nguyen, "Reconstruction and time reversal in thermoacoustic tomography in acoustically homogeneous and inhomogeneous media", *Inverse Problems,* vol.24, no.5, 055006, 2008.

[23] M. Xu, and L. V. Wang, "Universal back-projection algorithm for photoacoustic computed tomography", *Physical Review E,* vol.71, no.1, 016706, 2005.

[24] H. Chen, et al., "LEARN: Learned experts' assessment-based reconstruction network for sparse-data CT", *IEEE transactions on medical imaging,* vol.37, no.6, pp.1333-1347, 2018.

[25] Z. Jiang, et al., "Augmentation of CBCT reconstructed from under-sampled projections using deep learning", *IEEE transactions on medical imaging,* vol.38, no.11, pp.2705-2715, 2019.

[26] Z. Jiang, et al, "3D in vivo dose verification in prostate proton therapy with deep learning-based proton-acoustic imaging", *Physics in Medicine & Biology,* vol.67, no.21, 215012, 2022.

[27] B. Zhu, J. Z. Liu, S. F. Cauley, B. R. Rosen, and M. S. Rosen, "Image reconstruction by domain-transform manifold learning", *Nature,* vol.555, no.7697, pp.487–492, 2018.

[28] I. Häggström, C. R. Schmidtlein, G. Campanella, and T. J. Fuchs, "DeepPET: A deep encoder-decoder network for directly solving the PET image reconstruction inverse problem", *Med. Imag. Anal.,* vol.54, pp.253–262, 2019.

[29] J. Zhang, et al., "Ultrasound image reconstruction from plane wave radio-frequency data by self-supervised deep neural network", *Medical Image Analysis,* vol.70, 102018, 2021.

[30] Y. Luo, et al., "3D transformer-GAN for high-quality PET reconstruction", *International Conference on Medical Image Computing and Computer-Assisted Intervention,* pp.276-285, 2021.

[31] H Lan, D Jiang, C Yang, F Gao, and F Gao, "Y-Net: Hybrid deep learning image reconstruction for photoacoustic tomography in vivo", *Photoacoustics* vol.20, 100197, 2020.

[32] C Chen, Y Xing, H Gao, L Zhang, Z. Chen, "Sam's Net: A Self-Augmented Multi-Stage Deep-Learning Network for End-to-End Reconstruction of Limited Angle CT", *IEEE Transactions on*

*Medical Imaging*, vol.41, no.10, pp.2912-2924, 2022.

[33] N. Parmar, et al., "Image transformer", *International conference on machine learning,* pp.4055-4064, 2018.

[34] C. Matsoukas, J. F. Haslum, M. Söderberg, and K. Smith, "Is it time to replace cnns with transformers for medical images?", arXiv preprint arXiv:2108.09038, 2021.

[35] Liu, Ze, et al., "Swin transformer: Hierarchical vision transformer using shifted windows", *Proceedings of the IEEE/CVF International Conference on Computer Vision,* pp.10012-10022, 2021.

[36] J. Huang, Y. Wu, H. Wu, G. Yang, "Fast MRI Reconstruction: How Powerful Transformers Are?", arXiv preprint arXiv:2201.09400, pp. 2066-2070, 2022.

[37] B. E. Treeby, and B. T. Cox, "k-Wave: MATLAB toolbox for the simulation and reconstruction of photoacoustic wave fields", *Journal of biomedical optics*, vol.15, no.2, pp.021314-021314, 2010.

[38] M. Ahmad, L. Xiang, S. Yousefi, L. Xing, "Theoretical detection threshold of the proton-acoustic range verification technique", *Medical physics*, vol.42, no.10, pp.5735-5744, 2015.

[39] B. Carlier, et al., "Proton range verification with ultrasound imaging using injectable radiation sensitive nanodroplets: a feasibility study", *Physics in Medicine & Biology*, vol.65, no.6, 065013, 2020.

[40] J. Pietsch, et al., "Automatic detection and classification of treatment deviations in proton therapy from realistically simulated prompt gamma imaging data", *Medical Physics*, vol.50, no.1, pp.506-517, 2023.

[41] Y. Yu, Q. Pengyuan and P. Hao, "Feasibility study of 3D time-reversal reconstruction of proton-induced acoustic signals for dose verification in the head and the liver: A simulation study", *Medical Physics,* vol.48, no.8, pp.4485-4497, 2021.